\documentclass{Pos}

\title{Complex mass renormalization in EFT}

\ShortTitle{Complex mass renormalization}

\author{D.~Djukanovic\\
        Institut f\"ur Kernphysik, Johannes
Gutenberg-Universit\"at, J.J. Becherweg 45, \\ D-55099 Mainz,
Germany\\
        E-mail: \email{dalibor@kph.uni-mainz.de}}

\author{\speaker{J.~Gegelia}\\
        Institut f\"ur Kernphysik, Johannes
Gutenberg-Universit\"at, J.J. Becherweg 45, \\ D-55099 Mainz,
Germany \\ and \\ High Energy Physics Institute of TSU, Tbilisi,
Georgia\\
        E-mail: \email{gegelia@kph.uni-mainz.de}}

\author{A.~Keller\\
        Institut f\"ur Kernphysik, Johannes
Gutenberg-Universit\"at, J.J. Becherweg 45, \\ D-55099 Mainz,
Germany\\
        E-mail: \email{keller@kph.uni-mainz.de}}

\author{S.~Scherer\\
        Institut f\"ur Kernphysik, Johannes
Gutenberg-Universit\"at, J.J. Becherweg 45, \\ D-55099 Mainz,
Germany\\
        E-mail: \email{scherer@kph.uni-mainz.de}}

\abstract{We consider an effective field theory of unstable
particles (resonances) using the complex-mass renormalization. As an
application we calculate the masses and the widths of the $\rho$
meson and the Roper resonance.}

\FullConference{6th International Workshop on Chiral Dynamics, CD09\\
        July 6-10, 2009\\
        Bern, Switzerland}

\begin{document}

\section{Introduction}

   The construction of consistent chiral effective field theories with heavy degrees of freedom
is a non-trivial problem.
   For example, in baryon chiral perturbation theory the usual
power counting is violated if one uses the dimensional
regularization and the minimal subtraction scheme
\cite{Gasser:1987rb}.
   The current solutions to this problem either involve the heavy-baryon approach
\cite{Jenkins:1990jv} or use a suitably chosen renormalization
condition
\cite{Tang:1996ca,Becher:1999he,Gegelia:1999gf,Fuchs:2003qc}.
   Due to the small mass difference between the nucleon and the
$\Delta(1232)$ in comparison with the nucleon mass, the $\Delta$
resonance can be consistently included in the framework of effective
field theory
\cite{Hemmert:1997ye,Pascalutsa:2002pi,Bernard:2003xf,Hacker:2005fh,Pascalutsa:2006up}.

   On the other hand, the treatment of the $\rho$ meson or the inclusion of heavier baryon
resonances such as the Roper resonance is more complicated.
   We address the issue of power counting in such effective theories by using the
complex-mass renormalization scheme
\cite{Stuart:1990,Denner:1999gp,Denner:2006ic}, which can be
understood as an extension of the on-mass-shell renormalization
scheme to unstable particles.

   As an application we consider the masses and
the widths of the $\rho$ meson and the Roper resonance. More details
can be found in Refs.~\cite{Djukanovic:2009zn,Djukanovic:2009gt}.

\medskip

\section{Rho meson}

We start with the most general effective Lagrangian for $\rho$ and
$\omega$ mesons and pions in the parametrization of the model III of
Ref.~\cite{Ecker:1989yg}:
\begin{displaymath}
{\cal L}={\cal L}_{\pi}^{(2)}+{\cal L}_{\rho\pi}+{\cal L}_\omega
+{\cal L}_{\omega\rho\pi}+\cdots.
\end{displaymath}
   The individual expressions relevant for the calculations of
this work read
\begin{eqnarray}
{\cal L}_\pi^{(2)} & = & \frac{F^2}{4}\,{\rm Tr} \left[\partial_\mu
U \left(\partial^\mu
U\right)^\dagger\right]+\frac{F^2\,M^2}{4}\,{\rm Tr} \left(
U^\dagger+U\right), \nonumber\\
{\cal L}_{\rho\pi}&=& - \frac{1}{2}\,{\rm
Tr}\left(\rho_{\mu\nu}\rho^{\mu\nu}\right) + \left[ M_{\rho}^2 +
\frac{c_{x}\,M^2\,{\rm Tr} \left( U^\dagger+U\right) }{4}\right]
{\rm Tr}\left[\left( \rho^\mu-\frac{i\,\Gamma^\mu}{g}\right)\left(
\rho_{\mu}-\frac{i\,\Gamma_\mu}{g} \right)\right],\nonumber\\
{\cal L}_\omega&=&
 -\frac{1}{4}\left(
\partial_\mu\omega_{\nu}-\partial_\nu\omega_{\mu}\right)\left(
\partial^\mu\omega^\nu-\partial^\nu\omega^\mu\right)+\frac{M_{\omega}^2\,
\omega_{\mu}\omega^\mu}{2},\nonumber\\
{\cal L}_{\omega\rho\pi}&=&\frac{1}{2}\,
g_{\omega\rho\pi}\,\epsilon_{\mu\nu\alpha\beta}\, \omega^{\nu}\,
{\rm Tr}\left(\rho^{\alpha\beta} u^\mu
\right),\label{finallagrangian}
\end{eqnarray}
where
\begin{eqnarray}
U&=&u^2={\rm exp}\left(\frac{i\vec{\tau}\cdot\vec{\pi}}{F}\right), \
\ \ \
\rho^\mu  =  \frac{\vec\tau\cdot\vec\rho\,^\mu}{2},\nonumber\\
\rho^{\mu\nu} & = &
\partial^\mu\rho^\nu-\partial^\nu\rho^\mu - i
g\left[\rho^\mu,\rho^\nu\right] \,,\nonumber\\
\Gamma_\mu &= & \frac{1}{2}\,\bigl[ u^\dagger\partial_\mu u+u
\partial_\mu u^\dagger
u_\mu = i \left[ u^\dagger \partial_\mu u-u \partial_\mu u^\dagger
\right]. \label{somedefinitions}
\end{eqnarray}
All the fields and parameters in Eqs.~(\ref{finallagrangian}) are
bare quantities.
   In order to increase the readability of the expressions we have omitted
the usual subscript 0.
   In Eqs.~(\ref{finallagrangian}), $F$ denotes the pion-decay constant in the chiral
limit, $M^2$ is the lowest-order expression for the squared pion
mass, $M_\rho$ and $M_\omega$ refer to the bare $\rho$ and $\omega$
masses, $g$, $c_x$, and $g_{\omega\rho\pi}$ are coupling constants.
We use the KSFR relation \cite{Kawarabayashi:1966kd,Riazuddin:sw}
\begin{eqnarray}
M_\rho^2 & = & 2\,g^2 F^2 \,.\label{M0}
\end{eqnarray}

   To perform the renormalization we express the bare quantities
in terms of renormalized ones:
\begin{eqnarray}
M_{\rho,0} & = & M_R + \delta M_R \,,
\nonumber\\
c_{x,0}  & = & c_x+\delta c_x\,,\nonumber\\
 & \cdots & \label{renormparameters}
\end{eqnarray}
We apply the complex-mass renormalization scheme
\cite{Stuart:1990,Denner:1999gp,Denner:2006ic} and choose
$M_R^2=(M_\chi - i\, \Gamma_\chi/2)^2$ as the pole of the
$\rho$-meson propagator in the chiral limit.
   $M_\chi$ and $\Gamma_\chi$ are the pole mass and the width of the $\rho$ meson
in the chiral limit, respectively.
   Both are input parameters in our approach.
   In the complex-mass renormalization scheme, the counterterms are in general complex
quantities.

   The presence of large external momenta of the $\rho$ meson leads to a considerable
complication in the power counting for loop diagrams.
   It is necessary to investigate all possible flows of the external momenta through the
internal lines of a loop diagram.
   Next, one needs to determine
the chiral orders for all flows of external momenta.
   Finally, the smallest of these orders is defined as the chiral order of the given diagram.

   The power counting rules are as follows. Let $q$ collectively stand for a small quantity
such as the pion mass.
   A pion propagator counts as ${\cal O}(q^{-2})$ if it does not
carry large external momenta and as ${\cal O}(q^{0})$ if it does.
   A vector-meson propagator counts as ${\cal O}(q^{0})$
if it does not carry large external momenta and as ${\cal
O}(q^{-1})$ if it does.
   The pion mass counts as ${\cal O}(q^{1})$, the vector-meson mass as ${\cal O}(q^{0})$,
and the width as ${\cal O}(q^{1})$.
   Vertices generated by the effective Lagrangian of Goldstone bosons ${\cal
L}_\pi^{(n)}$ count as ${\cal O}(q^n)$.
   Derivatives acting on heavy vector mesons  count as ${\cal O}(q^0)$. The contributions of vector meson loops
can be absorbed systematically in the parameters of the effective
Lagrangian.

The dressed propagator, expressed in terms of the self energy
\begin{equation}
i\,\Pi^{ab}_{\mu\nu}(p)=i\,\delta^{ab}\,\left[ g_{\mu\nu}\Pi_{1}
(p^2)+p_\mu p_\nu\,\Pi_2 (p^2)\right] \label{VSEpar}
\end{equation}
has the form
\begin{equation}
i\, S^{a b}_{\mu\nu}(p) = -i \,\delta^{ab}\,\frac{g_{\mu\nu}-p_\mu
p_\nu\frac{1+ \Pi_2(p^2)}{M_R^2+\Pi_1(p^2)+p^2 \Pi_2(p^2)}}{p^2 -
M_R^2-\Pi_1(p^2)+i\,0^+}\,.\label{dressedprop}
\end{equation}
   The pole of the propagator is found as the (complex) solution to
the following equation:
\begin{equation}
z- M_R^2-\Pi_1(z)=0\,. \label{poleequation1}
\end{equation}
We define the pole mass and the width of the $\rho$ meson by
parameterizing
\begin{equation}
z=(M_\rho -i\,\Gamma/2)^2\,. \label{defmassandwidth}
\end{equation}
   The solution to Eq.~(\ref{poleequation1}) can be found
perturbatively as a loop expansion
\begin{equation}
z=z^{(0)}+ z^{(1)}+z^{(2)}+\cdots\,.\label{Mpoleloopexpansion}
\end{equation}
   Each of these terms has its own chiral expansion.
   Up to third chiral order the pole reads
\begin{equation}
z=z^{(0)}= M_R^2+c_x M^2\,. \label{poleequation3}
\end{equation}

\begin{figure}
\begin{center}
\includegraphics{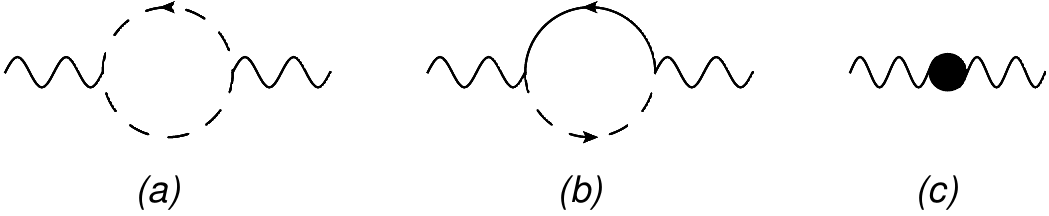}
\end{center}
\caption[]{\label{mnloops:fig} One-loop contributions to the
$\rho$-meson self-energy at ${\cal O}(q^3)$. The dashed, solid, and
wiggly lines correspond to the pion, the $\omega$ meson, and the
$\rho$ meson, respectively.}
\end{figure}

   The one-loop contributions to the vector self-energy up to ${\cal O}(q^3)$ are shown in
Fig.~\ref{mnloops:fig}.
   The contributions of diagrams (a) and (b) to $\Pi_1$
are given by
\begin{eqnarray}
\Pi_{1(a)} & = & \frac{g^2   \left[2 \,A_0(M^2)-\left(p ^2-4
M^2\right) \,B_0(p^2,M^2,M^2)\right]}{16\,\pi^2 (n-1)}\,, \nonumber\label{diag24} \\
\Pi_{1(b)} & = & - \frac{(n-2)\,g_{\omega \rho \pi}^2}{64\,\pi^2
(n-1)}\,\biggl\{ M^4\,B_0(p^2,M^2,M_\omega^2)
 -\left[2\,B_0(p^2,M^2,M_\omega^2)
M_\omega^2+A_0(M^2)-A_0(M_\omega^2)\right.\nonumber\\
&&\left. +2\,B_0(p^2,M^2,M_\omega^2) p^{2}\right] M^2
+B_0(p^2,M^2,M_\omega^2) p^{2 }+M_\omega^2
\left[B_0(p^2,M^2,M_\omega^2)
M_\omega^2\right.\nonumber\\
&&\left.+A_0(M^2)-A_0(M_\omega^2)\right]
-\left[2\,B_0(p^2,M^2,M_\omega^2)
M_\omega^2+A_0(M^2)+A_0(M_\omega^2)\right] p^{2}\biggr\}\,.
\label{diag25}
\end{eqnarray}
   Using dimensional regularization with $n$ space-time
dimensions, the loop functions read
\begin{eqnarray}
A_0\left(m^2\right) &=& 
-32 \pi ^2 \lambda\,  m^2-2\,m^2 \ln \frac{m}{\mu}\,, \nonumber\\
B_0\left(p^2,m_1^2,m_2^2\right) &=&
-32 \pi ^2 \lambda +2 \ln \frac{\mu}{m_2}-1\nonumber\\
&&  -\frac{1}
{2}\left(1+\frac{m_2^2}{m_1^2(\omega-1)}\right)\,{}_2F_1\left(1,2;3;1+\frac{m_2^2}{m_1^2(\omega-1)}\right)
 -  \frac{\omega}{2} \ {}_2F_1\left(1,2;3;\omega\right)\,,\nonumber\\
\omega & = & \frac{m_1^2
-m_2^2+p^2+\sqrt{\left(m_1^2-m_2^2+p^2\right)^2 -4m_1^2 p^2}}{2
m_1^2}\,, \label{oneandtwoPF}
\end{eqnarray}
where ${}_2F_1\left(a,b;c;z\right)$ is the standard hypergeometric
function, $\mu$ is the scale parameter and
\begin{equation}
\lambda =
\frac{1}{16\,\pi^2}\left\{\frac{1}{n-4}-\frac{1}{2}\,\left[\ln(4
\pi)+\Gamma'(1)+1\right]\right\}\,. \label{lambdadef}
\end{equation}

The $\rho\pi\pi$ vertex
in diagram (a) should count as ${\cal O}(q^0)$.
   However, its large component does not contribute to $\Pi_1$.
   Therefore, the  $\Pi_1$  part of diagram (a) has order
${\cal O}(q^4)$. Diagram (c) contains the contributions of the
counterterms.

In diagram (b) we take $M_\omega = M_R$.

We fix the counterterms such that the pole in the chiral limit stays
at $M_R^2$. The contributions  of diagrams (a), (b), and (c) to the
pole, expanded up to ${\cal O}(q^4)$, read
\begin{equation}
z^{(1)}=\frac{g^2M^4 }{16\,\pi^2 \,M_R^2}\left(3 -2\, \ln
\frac{M^2}{M_\chi^2}-2\,i\,\pi \right) -\frac{g_{\omega\rho\pi}^2
M^3 M_\chi}{24 \pi} -\frac{g_{\omega\rho\pi}^2 M^4 \left(\ln
\frac{M^2}{M_\chi^2}-1\right)}{32 \pi ^2} + \frac{i g_{\omega \rho
\pi}^2 \,M^3 \Gamma_\chi }{48 \pi }\,. \label{deltaM1}
\end{equation}

\section{Roper Resonance}

   The most general effective Lagrangian,
relevant for the subsequent calculation of the pole of the Roper
propagator at order ${\cal O}(q^3)$ reads:
\begin{equation}
\mathcal{L} =  \mathcal{L}_{0}+ {\cal L}_\pi^{(2)}+ \mathcal{L}_{R}
+\mathcal{L}_{N R}+\mathcal{L}_{\Delta R}\,, \label{lagrFull}
\end{equation}
where $\mathcal{L}_{0}$ is given by
\begin{eqnarray}
\mathcal{L}_{0} & = & \bar{N}\,(i
\partial\hspace{-.55em}/\hspace{.1em}-m_{N 0})N + \bar{R}(i
\partial\hspace{-.55em}/\hspace{.1em}-m_{R 0}) R \nonumber\\
&&
 - \bar\Psi_\mu\xi^{\frac{3}{2}}
\biggl[(i {\partial\hspace{-.55em}/\hspace{.1em}}-m_{\Delta
0})\,g^{\mu\nu}
-i\,(\gamma^{\mu}\partial^{\nu}+\gamma^{\nu}\partial^{\mu})+i\,\gamma^{\mu}
{\partial\hspace{-.55em}/\hspace{.1em}}\gamma^{\nu} + m_{\Delta
0}\,\gamma^{\mu}\gamma^{\nu}\biggr]\xi^{\frac{3}{2}} \Psi_\nu .
\end{eqnarray}
   Here, $N$ and $R$ denote nucleon and Roper isospin doublets 
with bare masses $m_{N 0}$ and $m_{R 0}$, respectively.
   $\Psi_\nu$ are the vector-spinor isovector-isospinor
Rarita-Schwinger fields of the $\Delta$ resonance
\cite{Rarita:1941mf} with bare mass $m_{\Delta 0}$ and
$\xi^{\frac{3}{2}}$ is the isospin-$3/2$ projector (see
Ref.~\cite{Hacker:2005fh} for more details).

   The interaction terms $\mathcal{L}_{R}$, $\mathcal{L}_{NR}$, and $\mathcal{L}_{\Delta R}$
are constructed following Ref.~\cite{Borasoy:2006fk}. At leading
order
\begin{equation}
{\cal L}_R^{(1)} = \frac{g_R}{2}\,\bar R \gamma^\mu\gamma_5 u_\mu
R\,. \label{Rpiint1}
\end{equation}
   The next-to-leading-order Roper Lagrangian is given by
\begin{equation}
{\cal L}_R^{(2)} = c_{1,0}^* \langle \chi_+\rangle\,\bar R\,R\,,
\label{Rpiint2}
\end{equation}
where $c_{1,0}^*$ is a coupling constant and $\chi_+ = M^2
(U+U^\dagger)$.
   The nucleon-Roper interaction reads
\begin{equation}
{\cal L}_{N R}^{(1)} = \frac{g_{N R}}{2}\,\bar R \gamma^\mu\gamma_5
u_\mu N+ {\rm h.c.}\,. \label{Rpiint11}
\end{equation}
   Finally, the leading-order interaction between the delta and the Roper
is given by
\begin{equation}
{\cal L}_{\Delta R}^{(1)}= - g_{\Delta R} \,\bar{\Psi}_{\mu}
\,\xi^{\frac{3}{2}} \,(g^{\mu\nu}
+\tilde{z}\,\gamma^{\mu}\gamma^{\nu})\, u_{\nu}\, R + {\rm h.c.}\,,
\label{pND}
\end{equation}
where we take the ''off-mass-shell parameter'' $\tilde z=-1$.

   To renormalize the loop diagrams, we apply the complex-mass
renormalization and write:
\begin{eqnarray}
m_{R 0} & = & z_\chi+\delta z_\chi\,,\nonumber \\
m_{N 0} & = & m+\delta m \,,\nonumber\\
m_{\Delta 0} & = &
z_{\Delta \chi} + \delta z_{\Delta \chi} \,,\nonumber\\
c_{1,0}^* & = & c_{1}^*+\delta c_{1}^*\,,\nonumber\\
&\cdots & \,, \label{barerensplit}
\end{eqnarray}
where $z_{\chi}$ is the complex pole of the Roper propagator in the
chiral limit, $m$ is the mass of the nucleon in the chiral limit,
and $z_{\Delta \chi}$ is the pole of the delta propagator in the
chiral limit.

\medskip

   We organize our perturbative calculation by applying the standard
power counting of Refs.~\cite{Weinberg:1991um}, i.e., an interaction
vertex obtained from an ${\cal O}(q^n)$ Lagrangian counts as order
$q^n$, a pion propagator as order $q^{-2}$, a nucleon propagator as
order $q^{-1}$, and the integration of a loop as order $q^4$.
   In addition, we assign the order $q^{-1}$ to the
$\Delta$ propagator and to the Roper propagator (carrying loop
momenta).
   Within the complex-mass renormalization scheme, such a power counting is
respected in the range of energies close to the Roper mass.

   The dressed propagator of the Roper
\begin{equation}
i S_R(p) = \frac{i}{p\hspace{-.45 em}/\hspace{.1em}-z_\chi
-\Sigma_R(p\hspace{-.45 em}/\hspace{.1em})}\,,\label{dressedDpr}
\end{equation}
where $\Sigma_R (p\hspace{-.45 em}/\hspace{.1em})$ denotes the
self-energy, has a complex pole which is obtained from the equation
\begin{equation}
z - z_\chi -\Sigma_R(z)=0\,. \label{poleequation}
\end{equation}

\medskip

\begin{figure}
\begin{center}
\includegraphics{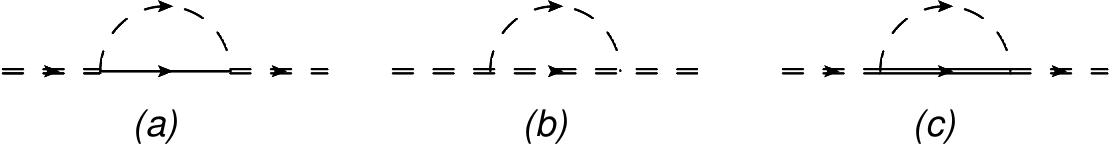}
\end{center}
\caption[]{\label{DeltaMassInd:fig} One-loop self-energy diagrams of
the Roper. The dashed, solid, double-dashed, and double-solid lines
correspond to the pion, nucleon, Roper, and delta, respectively.}
\end{figure}

   To order ${\cal O}(q^3)$ the Roper self-energy consists of a
tree-order contribution
\begin{equation}
\Sigma_{\rm tree} =  4\,c_1^*M^2 \,, \label{setree}
\end{equation}
and the loop diagrams shown in Fig.~\ref{DeltaMassInd:fig}.
   For the diagrams (a), (b), and (c) of Fig.~\ref{DeltaMassInd:fig}
we obtain
\begin{eqnarray}
\Sigma_{(a)} & = & \frac{3 g_{N R}^2}{128 \pi^2 F^2}
\left[\hat{O}_1(m) A_0\left(m^2\right)
+\hat{O}_2(m)A_0\left(M^2\right)
+\hat{O}_3(m) B_0\left(p^2,m^2,M^2\right)\right], \label{sigma1}\\
\Sigma_{(b)} & = & \frac{3 g_R^2}{128 \pi^2 F^2 }
\left[\hat{O}_1(z_\chi) A_0\left(z_\chi^2\right)
+\hat{O}_2(z_\chi)A_0\left(M^2\right)
+\hat{O}_3(z_\chi)B_0\left(p^2,z_\chi^2,M^2\right)\right],
\label{sigma2}\\
\Sigma_{(c)} & = & \frac{g_{\Delta  R}^2}{48 \pi^2 F^2 }
\left[\hat{O}_4 +\hat{O}_5 A_0\left(z_{\Delta \chi}^2\right)
+\hat{O}_6 A_0\left(M^2\right) +\hat{O}_7B_0\left(p^2,z_{\Delta
\chi}^2,M^2\right)\right], \label{sigma3}
\end{eqnarray}
where
\begin{eqnarray*}
\hat{O}_1(x)&=&p\hspace{-.45 em}/\hspace{.1em}\,\left(1+
\frac{x^2}{p^2}\right)
+2x,\\
\hat{O}_2(x)&=&p\hspace{-.45 em}/\hspace{.1em}
\left(1-\frac{x^2}{p^2}\right),\\
\hat{O}_3(x)&=& p\hspace{-.45
em}/\hspace{.1em}\left[-p^2\left(1-\frac{x^2}{p^2}\right)^2
+M^2\left(1+\frac{x^2}{p^2}\right)\right]+2M^2 x.\\
\hat{O}_4&=&\frac{1}{6}\left[3 p\hspace{-.45 em}/\hspace{.1em}
z_{\Delta \chi}^2-12 p^2 z_{\Delta \chi}-4 p\hspace{-.45
em}/\hspace{.1em} p^2  + 4 p^2 \frac{p^2-3M^2}{z_{\Delta
\chi}}+p\hspace{-.45 em}/\hspace{.1em}
\frac{2(p^2)^2-3 M^4-8 p^2 M^2}{z_{\Delta \chi}^2}\right],\\
\hat{O}_5&=&\frac{1}{p^2} \left[p\hspace{-.45 em}/\hspace{.1em}
z_{\Delta \chi}^2+2 p^2 z_{\Delta \chi} -p\hspace{-.45
em}/\hspace{.1em} \left(2 M^2+p^2\right) +2 p^2
\frac{p^2-M^2}{z_{\Delta \chi}} +p\hspace{-.45 em}/\hspace{.1em}
\frac{\left(M^2-p^2\right)^2}{z_{\Delta\chi}^2}\right],\\
\hat{O}_6&=&-\frac{1}{p^2}\left[p\hspace{-.45 em}/\hspace{.1em}
z_{\Delta \chi}^2+2 p^2 z_{\Delta \chi}-2 M^2 p\hspace{-.45
em}/\hspace{.1em} -2 p^2 \frac{M^2+p^2}{z_{\Delta \chi}}
+p\hspace{-.45em}/\hspace{.1em}\frac{M^4-3 p^2
M^2-(p^2)^2}{z_{\Delta\chi}^2}\right],\\
\hat{O}_7&=&-\frac{1}{p^2}\left[p\hspace{-.45 em}/\hspace{.1em}
z_{\Delta \chi}^2+2 p^2 z_{\Delta \chi} +p\hspace{-.45
em}/\hspace{.1em} \left(p^2-M^2\right)\right]
 \left[z_{\Delta \chi}^2-2 \left(M^2+p^2\right)
+\frac{\left(M^2-p^2\right)^2}{z_{\Delta\chi}^2}\right].
\end{eqnarray*}

   To implement the complex-mass renormalization scheme, in analogy
to Ref.~\cite{Fuchs:2003qc}, we expand the self-energy loop diagrams
in powers of $M$, $p\hspace{-.45 em}/\hspace{.1em} -z_\chi$, and
$p^2-z_\chi^2$, which all count as ${\cal O}(q)$.
   We subtract those terms which violate
the power counting.
   The subtraction terms at
$p\hspace{-.45 em}/\hspace{.1em} =z_\chi$ read
\begin{eqnarray}
\Sigma_{(a)}^{\rm ST} & = &  -\frac{3 g_{N R}^2 (m+z_\chi)^2}{128
\pi^2 F^2 z_\chi} \biggl[(m-z_\chi)^2
B_0\left(z_\chi^2,0,m^2\right)-A_0\left(m^2\right)\biggr]\nonumber\\
&& + \frac{3 g_{N R}^2 (m+z_\chi) M^2}{64 \pi^2 F^2 z_\chi^3 }
\left[-2 m^3\,\ln\frac{m}{\mu} -i \pi m^3+z_\chi^2 m-32 \pi^2
z_\chi^3  \lambda \right.\nonumber\\
&&\left. +\left(m^3-z_\chi^3\right) \ln\frac{z_\chi^2-m^2}{\mu^2} +i
\pi z_\chi^3
\right],\nonumber \label{STsigma1}\\
\Sigma_{(b)}^{\rm ST} & = &  \frac{3 g_R^2 z_\chi}{32 \pi^2 F^2}\
A_0\left(z_\chi^2\right) -\frac{3 g_R^2 z_\chi M^2}{32 \pi^2 F^2}
\left[32 \pi ^2 \lambda +2 \ln \frac{z_\chi}{\mu}-1\right]\nonumber \label{STsigma2},\\
\Sigma_{(c)}^{\rm ST} & = & -\frac{g_{\Delta  R}^2}{288 F^2\pi^2
z_{\Delta \chi}^2 z_\chi} \left[ 6 (z_{\Delta \chi}-z_\chi)^2
(z_{\Delta \chi}+z_\chi)^4B_0\left(z_\chi^2,0,z_{\Delta
\chi}^2\right)
\right.\nonumber\\
&&+z_\chi^2 \bigl(-3 z_{\Delta \chi}^4 +12 z_\chi z_{\Delta \chi}^3
+4 z_\chi^2 z_{\Delta \chi}^2 -4 z_\chi^3 z_{\Delta \chi}
-2z_\chi^4\bigr)\nonumber\\
&&\left. -6 \left(z_{\Delta \chi}^4+2 z_\chi z_{\Delta
\chi}^3-z_\chi^2 z_{\Delta \chi}^2+2 z_\chi^3 z_{\Delta
\chi}+z_\chi^4\right)
A_0\left(z_{\Delta \chi}^2\right)\right]\nonumber\\
&& + \frac{g_{\Delta  R}^2 M^2}{72 \pi^2 F^2 z_{\Delta \chi}^2
z_\chi^3} \left[-6 i \pi z_{\Delta \chi}^6-6 (2 z_{\Delta \chi}+3
z_\chi) z_{\Delta \chi}^5\ln\frac{z_{\Delta \chi}}{\mu} -9 i \pi
z_\chi  z_{\Delta \chi}^5
+6 z_\chi^2 z_{\Delta \chi}^4\right. \nonumber\\
&& +9 z_\chi^3 z_{\Delta \chi}^3+3 z_\chi^4 z_{\Delta \chi}^2 -288
\pi ^2 \lambda z_\chi^5 z_{\Delta \chi} +9 i  \pi z_\chi^5 z_{\Delta
\chi} +z_\chi^6
-192 \pi^2\lambda z_\chi^6 \nonumber\\
&& \left.+\left(6 z_{\Delta \chi}^6+9 z_\chi z_{\Delta \chi}^5-9
z_\chi^5 z_{\Delta \chi}-6 z_\chi^6\right) \ln
\frac{z_\chi^2-z_{\Delta \chi}^2}{\mu^2} +6 i \pi z_\chi^6 \right].
\label{STsigma3}
\end{eqnarray}
   The above expressions of Eq.~(\ref{STsigma3}) are exactly canceled
by contributions of $\delta z_\chi$ and $\delta c_{1}^*$. The pole
of the Roper propagator to third order is given by the expression
\begin{equation}
z = z_\chi -4\,c_1^*M^2 +
\left[\Sigma_{(a)}+\Sigma_{(b)}+\Sigma_{(c)}\right]_{p\hspace{-.3
em}/\hspace{.1em}=z_\chi}- \Sigma_{(a)}^{\rm ST}-\Sigma_{(b)}^{\rm
ST}-\Sigma_{(c)}^{\rm ST}\,. \label{pole}
\end{equation}
   It is easily shown that the expansion of Eq.~(\ref{pole}) satisfies the power counting,
i.e.~is of ${\cal O}(q^3)$.

\medskip

\section{Summary}
    We have considered an effective field theory of
resonances interacting with Goldstone bosons using the complex-mass
renormalization scheme. A systematic power counting emerging within
this scheme allows one to calculate the physical quantities in
powers of small parameters. As an application we have calculated the
pole masses and the widths of the $\rho$ meson and the Roper
resonance which are of particular interest in the context of lattice
extrapolations.
   The masses and the widths in the chiral limit are considered
as input parameters within this approach.

\acknowledgments

This work was supported by the Deutsche Forschungsgemeinschaft (SFB
443). We thank the organizers for a very pleasant meeting.

\end{document}